# INTENSITY MOMENTS OF A LASER BEAM FORMED BY SUPERPOSITION OF HERMITE-GAUSSIAN MODES

A. Ya. Bekshaev

I. I. Mechnikov Odessa State University

SUMMARY

Expressions are obtained for the Wigner function moments of a paraxial light beam represented by arbitrary coherent superposition of Hermite-Gaussian beams with plane wave fronts. Possibilities are discussed for application of the obtained results to modeling real laser beams and to design of optical systems implementing prescribed transformations of a beam transverse structure.

# INTENSITY MOMENTS OF A LASER BEAM FORMED BY SUPERPOSITION OF HERMITE-GAUSSIAN MODES

A. Ya. Bekshaev

I. I. Mechnikov Odessa State University

The spatial-angular intensity distribution (SAID) of radiation is one of the most important characteristics of a light beam. Its detailed description generally requires specifying a great number of parameters, which is practically inconvenient and, quite often, superfluous. Therefore, various schemes of more rational SAID characterization have been proposed among which two approaches seem to be the most suitable and universal: a systematic use of the Wigner function moments [1, 2] and the beam representation as a superposition of the standard Hermite-Gaussian (HG) modes [3]. The first approach is distinguished by theoretical self-consistency as well as by descriptiveness and immediate measurability of relevant parameters, the second one is useful in calculations and provides possibility to establish direct relations between the SAID parameters and the physical picture of the laser beam generation. Obviously, it would be quite suitable to know the correspondence between the beam characteristics employed in both approaches, and the present report contributes to solution of this problem in important special cases. Results of the below analysis may be used, in particular, for practical representation of an arbitrary paraxial light beam by means of a certain superposition of HG modes. This will enhance the possibilities and the area of applicability of the "embedded" laser beam method [1,2], enhancing and supplementing the known recommendations for beam modeling by an incoherent mixture of two Gaussian beams or by an off-axis Gaussian beam [1].

As usual, we consider a beam propagating along axis $z$, and the local instantaneous optical field (e.g., electric field strength) is expressed through the slowly varying complex amplitude $u$ [3]

$$E(x,y,z,t) = \mathrm{Re}\left\{u(x,y,z)\exp\left[i(kz-\omega t)\right]\right\}$$

where $x$, $y$ are Cartesian coordinates in the beam cross section, $k$ and $\omega$ are the radiation wave number and frequency, $t$ stands for the time. For a superposition of HG modes, the normalized transverse distribution of the beam complex amplitude obtains the representation

$$u(x,y) = \sum_{m,n} a_{mn} u_{mn}(x,y), \quad \sum_{m,n} |a_{mn}|^2 = 1, \qquad (1)$$

with the distribution of a normalized HG mode [3]

$$u_{mn}(x,y) = \left(2^{m+n} \pi m! n! b_x b_y\right)^{-1/2} H_m\left(\frac{x}{b_x}\right) H_n\left(\frac{y}{b_y}\right) \exp\left[-\frac{1}{2}\left(\frac{x^2}{b_x^2} + \frac{y^2}{b_y^2}\right)\right]. \qquad (2)$$

Here $m \geq 0$, $n \geq 0$ are the integer mode indices, $H_m$ denotes the Hermite polynomial of $m$-th order, $b_x$ and $b_y$ are orthogonal transverse sizes of the Gaussian envelope [3]. Due to completeness of the set of functions (2), the expansion (1) exists for every function $u(x,y)$, i. e. for every real paraxial beam.

The moment matrix of a paraxial beam is defined by equality [1]

$$\mathbf{Q} = \begin{pmatrix} \mathbf{Q}_{11} & \mathbf{Q}_{12} \\ \mathbf{Q}_{21} & \mathbf{Q}_{22} \end{pmatrix} = 2k \int \begin{pmatrix} x^2 & xy & xp_x & xp_y \\ yx & y^2 & yp_x & yp_y \\ p_x x & p_x y & p_x^2 & p_x p_y \\ p_y x & p_y y & p_y p_x & p_y^2 \end{pmatrix} I(x,y,p_x,p_y) dx\, dy\, dp_x\, dp_y,$$

where $\mathbf{Q}_{11}$ ... $\mathbf{Q}_{22}$ are 2×2 matrices, $p_x$ and $p_y$ are arguments of the beam angular spectrum, $I(x,y,p_x,p_y)$ is the beam Wigner function [4] normalized by condition $\int I(x,y,p_x,p_y) dx\, dy\, dp_x\, dp_y = 1$; integrals are taken within infinite limits of each variable, while the presumed spatial confinement of the beam provides their convergence.[1]

---

[1] The equivalent representation of the moment matrix $\mathbf{M} = \begin{pmatrix} \mathbf{M}_{11} & \mathbf{M}_{12} \\ \mathbf{M}_{21} & \mathbf{M}_{22} \end{pmatrix} = \frac{\mathbf{Q}}{2k}$ is also often used (see, e.g., Refs. [10–12]).

Our task is to find the moment matrix for the beam described by Eqs. (1) and (2). Using formulas from Appendix A of paper [1] as well as recurrent relations for the Hermite polynomials [5], after rather cumbersome transformations we obtain the representation

$$Q = 2 \begin{pmatrix} kb_x^2(K_x + L'_x) & kb_x b_y(M' + N') & -L''_x & \frac{b_x}{b_y}(M'' - N'') \\ kb_x b_y(M' + N') & kb_y^2(K_y + L'_y) & -\frac{b_y}{b_x}(M'' + N'') & -L''_y \\ -L''_x & -\frac{b_y}{b_x}(M'' + N'') & \frac{1}{kb_x^2}(K_x - L'_x) & \frac{1}{kb_x b_y}(M' - N') \\ \frac{b_x}{b_y}(M'' - N'') & -L''_y & \frac{1}{kb_x b_y}(M' - N') & \frac{1}{kb_y^2}(K_y - L'_y) \end{pmatrix},$$

(3)

where

$$K_x = \frac{1}{2}\sum_{m,n}|a_{mn}|^2(2m+1), \quad K_y = \frac{1}{2}\sum_{m,n}|a_{mn}|^2(2n+1), \quad (4)$$

$$L_x = \sum_{m,n} a_{mn} a^*_{m+2,n}\sqrt{(m+1)(m+2)}, \quad L_y = \sum_{m,n} a_{mn} a^*_{m,n+2}\sqrt{(n+1)(n+2)}, \quad (5)$$

$$M = \sum_{m,n} a_{m,n+1} a^*_{m+1,n}\sqrt{(m+1)(n+1)}, \quad N = \sum_{m,n} a_{mn} a^*_{m+1,n+1}\sqrt{(m+1)(n+1)}, \quad (6)$$

one prime (two primes) denote the real (imaginary) part of a complex quantity.

Results (3) - (6) give a general ground for the solution of our problem. As could be expected, there is no one-to-one mapping between the two schemes of the beam characterization. For every HG mode superposition, one can determine the unique moment matrix but the inverse procedure – choice of a HG superposition with a given moment matrix – can be realized in multitude of ways (of course, if no explicit restrictions for coefficients $a_{mn}$ are imposed). One can easily see that for arbitrary moment matrix possessing not more than ten independent elements, from Eqs. (3) -

(6) the relations follow which unambiguously determine ten real quantities forming $K_x$, $K_y$, $L_x$, $L_y$, $M$ and $N$. Their knowledge allows to find, at most, ten coefficients $a_{mn}$, and, therefore, corresponding representation of expansion (1) can be unique only if number of the superposition members is limited or they obey certain interrelations.

Nevertheless, in spite of the mentioned ambiguity of modeling an arbitrary moment matrix by the superposition of the form (1), Eqs. (3) - (6) allow us to determine those parameters of HG modes which are especially important in this procedure. Now, with several examples, consider some consequences and ways of utilization of results (3) - (6).

1. Let the beam be represented by a "pure" HG mode (2); then in Eqs. (3) - (6) only one of coefficients $a_{mn}$ equals to unity while all the rest reduce to zeroes. The moment matrix becomes diagonal:

$$\mathbf{Q}_{mn} = \begin{pmatrix} \mathbf{Q}_{11} & \mathbf{Q}_{12} \\ \mathbf{Q}_{21} & \mathbf{Q}_{22} \end{pmatrix}_{mn} = \text{diag}\left[ kb_x^2(2m+1), kb_y^2(2n+1), \frac{2m+1}{b_x^2 k}, \frac{2n+1}{b_y^2 k} \right]. \quad (7)$$

Invariants of this moment matrix (see [1]) possess the form

$$\Lambda_x = \mathrm{M}_x^2 = 2m+1; \quad \Lambda_y = \mathrm{M}_y^2 = 2n+1, \quad (8)$$

$$\det \mathbf{Q}_{mn} = (2m+1)^2 (2n+1)^2, \quad \mathrm{Sp}\left(\mathbf{Q}_{11}\mathbf{Q}_{22} - \mathbf{Q}_{12}^2\right)_{mn} = (2m+1)^2 + (2n+1)^2. \quad (9)$$

Relations (8) and (9) permit one to find a HG beam with the given moment matrix. As is seen from the formulas, this is possible not always but only for beams whose moment matrix has integer invariants $\Lambda_x$ and $\Lambda_y$.

2. Now consider a beam with simple astigmatism [3], in which the transverse distribution of the complex amplitude is symmetric or antisymmetric with respect to the coordinate axes (this class embraces, for example, all axially symmetric beams that occur in practice most often). HG mode (2) is a special case of this beam; its symmetry or antisymmetry is indicated by evenness or oddness of the corresponding index. Obviously, superposition (1) preserves these properties if all its members have the same parity. Consequently, for beams with simple astigmatism $M = N = 0$ and the moment matrix, as should be expected, acquires the distinctive for such beams form with zeros at crossings of rows and columns with different parities. In this case, the beam is described by two 2×2 moment matrices $Q_x$ and $Q_y$ where, for example,

$$Q_x = 2k \int \begin{pmatrix} x^2 & xp_x \\ p_x x & p_x^2 \end{pmatrix} I(x, y, p_x, p_y) dx\, dy\, dp_x dp_y = 2 \begin{pmatrix} kb_x^2(K_x + L'_x) & -L''_x \\ -L''_x & \frac{1}{kb_x^2}(K_x - L'_x) \end{pmatrix}.$$

The beam invariants satisfy the relationships

$$\Lambda_x^2 = \det Q_x = 4\left(K_x^2 - |L_x|^2\right), \quad \Lambda_y^2 = \det Q_y = 4\left(K_y^2 - |L_y|^2\right),$$

which, for the pure HG mode, lead to equalities (8) and (9).

3. Yet another interesting situation appears when all coefficients of the series (1) are real. This means that phases of all the superposition members are equal, and since each of them describes a beam with plane wavefront, the resulting beam also possesses a plane front. Really, then all imaginary parts of amounts (4) - (6) equal to zeroes and matrix (3) turns out to be a block-diagonal matrix with $Q_{12} = Q_{21} = 0$, which, according to [1], is a characteristic sign of the beam waist [6] (a section with plane wavefront).

4. Typical cases of realization of the form (2) superpositions occur upon the multimode generation in laser resonators with degenerate modes of type (1). Let us restrict ourselves by the situation when $b_x = b_y = b$; then all beams (2) with identical values of $m + n$ are degenerated. We consider two simplest superpositions that belong to the case $m + n = 1$. Herewith, in expansion (1) only two coefficients $a_{mn}$ can differ from zero: $a_{10}$ and $a_{01}$.

a) Let $a_{10} = \frac{1}{\sqrt{2}}$, $a_{01} = \frac{i}{\sqrt{2}}$. Such a beam is described by the complex amplitude distribution

$$u_a(x, y) = \frac{1}{b^2\sqrt{\pi}}(x + iy)\exp\left(-\frac{x^2 + y^2}{2b^2}\right) = \frac{1}{b^2\sqrt{\pi}} r \exp\left(-\frac{r^2}{2b^2} + i\varphi\right) \quad (10)$$

($r$ and $\varphi$ are the polar coordinates in the beam cross section) and represents an elementary example of a singular beam with the screw wavefront dislocation[2] [7]. In this case, Eqs. (4) – (6) give $K_x = K_y = 1$, $L_x = L_y = 0$, $M = i/2$, $N = 0$ and the moment matrix obtains the form

$$\mathsf{Q} = \mathsf{Q}_a = \begin{pmatrix} 2kb^2\mathsf{I} & \mathsf{J} \\ -\mathsf{J} & \frac{2}{kb^2}\mathsf{I} \end{pmatrix}, \quad (11)$$

where $\mathsf{I}$ is the unity 2×2 matrix and $\mathsf{J} = \begin{pmatrix} 0 & 1 \\ -1 & 0 \end{pmatrix}$ is the simplest antisymmetric 2×2 matrix.

б) The next superposition differs from the previous one by the fact that both non-zero coefficients of the expansion (1) are real: $a_{10} = a_{01} = \frac{1}{\sqrt{2}}$. Such a "small" mathematical difference leads to noticeable physical consequences. The corresponding complex amplitude distribution

$$u_b(x, y) = \frac{1}{b^2\sqrt{\pi}}(x + y)\exp\left(-\frac{x^2 + y^2}{2b^2}\right) = \sqrt{\frac{2}{\pi}}\frac{x'}{b^2}\exp\left(-\frac{x'^2 + y'^2}{2b^2}\right) \quad (12)$$

represents a "pure" $HG_{10}$ mode in the coordinate frame turned around the axis $z$ by 45°: $x' = (x + y)/\sqrt{2}$, $y' = (-x + y)/\sqrt{2}$, and the moment matrix of beam (12) has a block-diagonal form

---

[2] Here a reference to singular optics and orbital angular momentum [13] would be quite relevant but I did not know about these notions when the paper was prepared.

$$Q = Q_b = \begin{pmatrix} kb^2 \mathsf{P} & \mathsf{0} \\ \mathsf{0} & \dfrac{1}{kb^2}\mathsf{P} \end{pmatrix}, \quad (13)$$

where $\mathsf{0}$ is zero 2×2 matrix, $\mathsf{P} = \begin{pmatrix} 2 & 1 \\ 1 & 2 \end{pmatrix}$.

This example is rather demonstrative since the difference in forms of matrices $Q_a$ and $Q_b$ can be directly attributed to the wave front singularity of the first beam. However, one can easily make sure that the invariants of matrices (11) and (13) are equal:

$$\det Q_{a,b} = 9, \; \mathrm{Sp}\left(Q_{11}Q_{22} - Q_{12}^2\right)_{a,b} = 10;$$

hence, matrix $Q_a$ can be transformed to form (13) by some symplectic transformation [1]. As both beams (10) and (12) are formed by combining the same HG modes, this means that the beam with transverse distribution (10) can be transformed to form (12) with the help of a certain optical system consisting only of quadratic phase correctors (parabolic lenses and mirrors, generally astigmatic) and transversely homogeneous (free) intervals [8].

This conclusion may be important for adaptive optics where the correction of phase aberrations causing wavefront dislocations is usually treated as a complicated problem due to impossibility to create a correcting mirror of the required shape [7,9]. As can be seen, in this case at least, a beam with the wavefront dislocation can be transformed into a beam with the same quality and without phase singularities, which, if necessary, can afterwards be corrected by usual means of adaptive optics.

In conclusion, we present an example of such a transformation. The algorithm of arbitrary moment matrix transformation to the block-diagonal form (Ref. [1], Sec. 4) enables to calculate an optical system after transmission through which the beam (10) accepts form (12). Such a system can only be determined ambiguously, and the practical choice of a certain option is ultimately dictated by considerations of convenience. In particular, one can show that

$$Q_b = H Q_a \tilde{H},$$

where

$$H = \frac{1}{\sqrt{2}} \begin{pmatrix} 1 & 0 & kb^2 & 0 \\ 0 & -1 & 0 & kb^2 \\ -(kb^2)^{-1} & 0 & 1 & 0 \\ 0 & -(kb^2)^{-1} & 0 & -1 \end{pmatrix}$$

is the transmission matrix (generalized *ABCD* matrix [1,2,6]) of the transforming optical system. This matrix has zero elements at crossings of rows and columns of different parities, i.e. describes an optical system with simple astigmatism. It can be decomposed into two independent submatrices

$$H_x \equiv \begin{pmatrix} A_x & B_x \\ C_x & D_x \end{pmatrix} = \frac{1}{\sqrt{2}} \begin{pmatrix} 1 & kb^2 \\ -(kb^2)^{-1} & 1 \end{pmatrix}, \quad H_y \equiv \begin{pmatrix} A_y & B_y \\ C_y & D_y \end{pmatrix} = \frac{1}{\sqrt{2}} \begin{pmatrix} -1 & kb^2 \\ -(kb^2)^{-1} & -1 \end{pmatrix}. \quad (14)$$

It is easy to understand the physical meaning of these matrices. Both of them admit the representation

$$H_{x,y} = \begin{pmatrix} -g_{2x,y} & z_{01} z_{02}/f \\ -1/f & -g_{1x,y} \end{pmatrix}$$

corresponding to the "waist to waist" transformation ([6], Sec. 8.1); here $f = \sqrt{2} kb^2$, $z_{01} = z_{02} = kb^2$ are coinciding confocal parameters (Raleigh ranges) of the input and output beams, $g_{1x} = g_{2x} = 1/\sqrt{2}$, $g_{1y} = g_{2y} = -1/\sqrt{2}$. We can readily imagine an optical system implementing the transformation H; one of possible options is presented in the figure.[3] This system is formed by two identical astigmatic lenses placed in the input and output reference planes. The distance between the lenses equals to $L = kb^2/\sqrt{2}$, focal lengths of the lenses in plane $xz$ amounts to $f_x = \sqrt{2}L\left(\sqrt{2}-1\right)^{-1}$, and, in plane $yz$, $f_y = \sqrt{2}L\left(\sqrt{2}+1\right)^{-1}$. This causes that in plane $yz$ the beam experiences stronger focusing than in plane $xz$, and the $HG_{01}$ and $HG_{10}$ components of (10) obtains different additional phase shifts[4] which compensate the initial phase difference between them. Really, when transmitting the system with matrices (14), mode $HG_{10}$ gets the multiplier [3] $\left(A_x + \dfrac{B_x}{\tilde{\rho}}\right)^{-3/2}\left(A_y + \dfrac{B_y}{\tilde{\rho}}\right)^{-1/2}$, and mode $HG_{01}$, in turn – the multiplier $\left(A_x + \dfrac{B_x}{\tilde{\rho}}\right)^{-1/2}\left(A_y + \dfrac{B_y}{\tilde{\rho}}\right)^{-3/2}$ (here $\tilde{\rho} = -ikb^2$ is the complex radius of the input beam wavefront curvature). It can be easily seen from (14) that the ratio of these factors exactly equals to $i$, and thus the phase difference between the HG components of beam (10) vanishes after transmitting the optical system.

---

[3] Obviously, this optical system is a variant of the known mode converter transforming between HG and Laguerre-Gaissian modes [14].

Considered examples, list of which can be continued, illustrate peculiarities of the two approaches to the beam characterization and demonstrate the efficacy of their combined application. All this will facilitate the utilization of the results obtained in specific investigations of laser beams.

---

[4] In the newer literature, this shift is commonly referred to as Gouy phase [13].

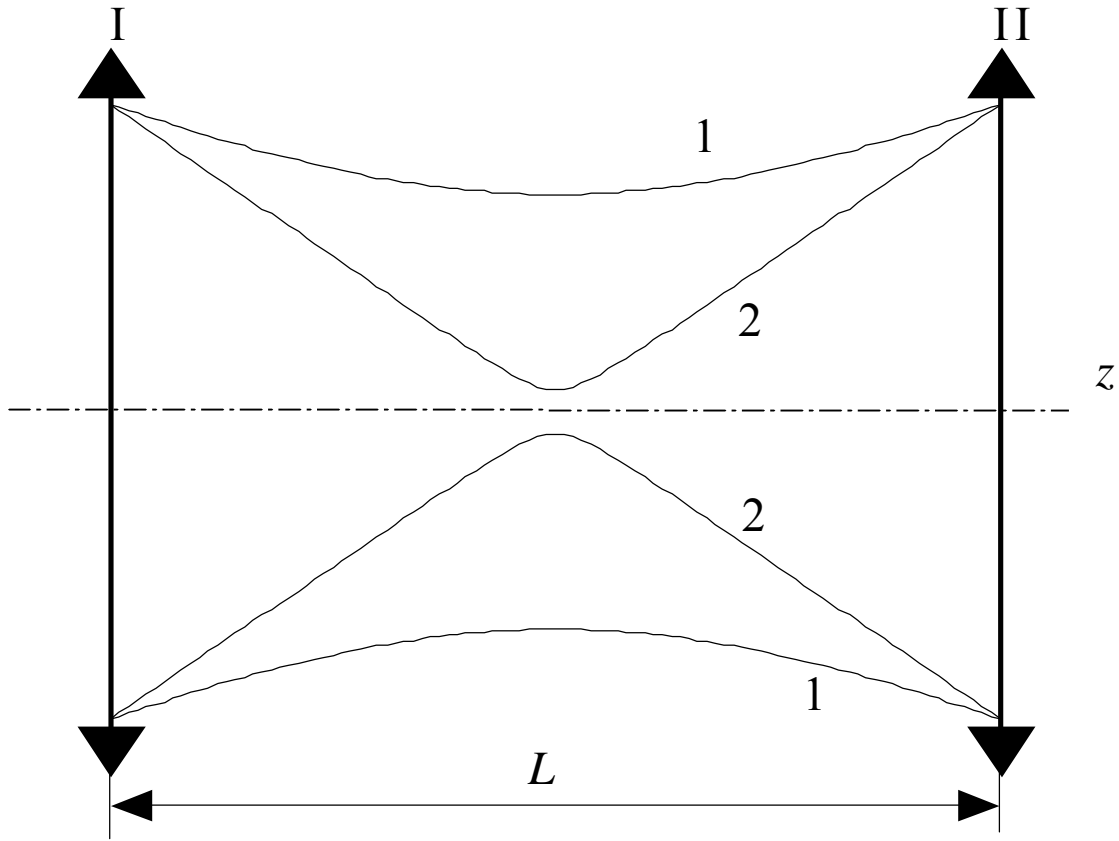

Figure caption

Optical system removing the wavefront dislocation: $z$ is the system axis; I, II are the input and output reference planes in which the thin astigmatic lenses are situated; 1 and 2 are schematic contours of the longitudinal beam sections in planes $xz$ and $yz$, correspondingly.